\documentstyle[12pt]{article}
\def\ie{{\em i.e.}}
\def\eg{{\em e.g.}}

\def\beq{\begin{equation}}
\def\eeq{\end{equation}}

\catcode`\@=11 

\def\vev#1{\left\langle #1\right\rangle}
\def\lsim{\mathrel{\mathpalette\@versim<}}
\def\gsim{\mathrel{\mathpalette\@versim>}}
\def\@versim#1#2{\vcenter{\offinterlineskip
    \ialign{$\m@th#1\hfil##\hfil$\crcr#2\crcr\sim\crcr } }}
\def\etal{{\em et. al.}}
\def\JL{J. L. Lopez}
\def\DVN{D. V. Nanopoulos}

\def\r#1{$\bf#1$}
\def\rb#1{$\bf\overline{#1}$}

\def\t1{{\tilde 1}}

\def\MeV{\,{\rm MeV}}
\def\GeV{\,{\rm GeV}}

\def\to{\rightarrow}

\def\NPB#1#2#3{Nucl. Phys. B {\bf#1} (19#2) #3}
\def\PLB#1#2#3{Phys. Lett. B {\bf#1} (19#2) #3}

\def\PRD#1#2#3{Phys. Rev. D {\bf#1} (19#2) #3}

\def\PRT#1#2#3{Phys. Rep. {\bf#1} (19#2) #3}

\def\TAMU#1{Texas A \& M University preprint CTP-TAMU-#1}

\begin{document}

\begin{flushright}
\baselineskip=12pt
{CERN-TH.7138/94}\\
{CERN-LAA/94-07}\\
{CTP-TAMU-05/94}\\
{ACT-31/94}\\
\end{flushright}

\begin{center}
\vglue 0.6cm
{\Large\bf The top-quark mass in SU(5)xU(1) supergravity\\}
\vspace{0.2cm}
\vglue 1cm
{JORGE L. LOPEZ$^{(a),(b)}$, D. V. NANOPOULOS$^{(a),(b),(c)}$,
and A. ZICHICHI$^{(d)}$\\}
\vglue 0.4cm
{\em $^{(a)}$Center for Theoretical Physics, Department of Physics, Texas A\&M
University\\}
{\em College Station, TX 77843--4242, USA\\}
{\em $^{(b)}$Astroparticle Physics Group, Houston Advanced Research Center
(HARC)\\}
{\em The Mitchell Campus, The Woodlands, TX 77381, USA\\}
{\em $^{(c)}$CERN, Theory Division, 1211 Geneva 23, Switzerland\\}
{\em $^{(d)}$CERN, 1211 Geneva 23, Switzerland\\}
\baselineskip=12pt

\vglue 1cm
{\tenrm ABSTRACT}
\end{center}
{\rightskip=3pc
 \leftskip=3pc
\noindent
We show that the currently experimentally preferred values of the top-quark
mass (\ie, $130\lsim m_t\lsim180\GeV$) are naturally understood in the context
of string models, where the top-quark Yukawa coupling at the string scale
is generically given by $\lambda_t={\cal O}(g)$, with $g$ the unified gauge
coupling. A detailed study of the Yukawa sector of $SU(5)\times U(1)$
supergravity shows that the ratio of the bottom-quark to tau-lepton Yukawa
couplings at the string scale is required to be in the range
$0.7\lsim\lambda_b/\lambda_\tau\lsim1$, depending on the values of $m_t$ and
$m_b$. This result is consistent with $SU(5)\times U(1)$ symmetry, which
does {\em not} require the equality of these Yukawa couplings in the unbroken
symmetry phase of the theory. As a means of possibly predicting the
value of $m_t$, we propose a procedure whereby the size of the allowed
parameter space is determined as a function of $m_t$, since all sparticle and
Higgs-boson masses and couplings depend non-trivially on $m_t$. At present,
no significant preference for particular values of $m_t$ in $SU(5)\times U(1)$
supergravity is observed, except that high-precision LEP data requires
$m_t\lsim180\GeV$.}
\vspace{1cm}
\begin{flushleft}
\baselineskip=12pt
{CERN-TH.7138/94}\\
{CERN-LAA/94-07}\\
{CTP-TAMU-05/94}\\
{ACT-31/94}\\
January 1994
\end{flushleft}
\vfill\eject
\setcounter{page}{1}
\pagestyle{plain}

\baselineskip=14pt

\section{Introduction}
The origin of elementary particle masses is one of the most profound questions
in physics. Modern field theories try to answer this question, in the context
of spontaneously broken gauge symmetries, through vacuum expectation values
(vevs) of elementary or composite scalar Higgs fields. In general the masses of
all particles (scalars, fermions, gauge bosons) are proportional to this (or
these) vev(s). The proportionality coefficients are: the quartic couplings
($\lambda$) for the scalars, the Yukawa couplings ($y$) for the fermions, and
the gauge couplings ($g$) for the gauge bosons. Thus, schematically we have:
\begin{eqnarray}
	m_s &=& \lambda^{1/2}\vev{\rm vev},\label{one}\\
	m_f &=& y \vev{\rm vev},\label{two}\\
	m_g &=& g \vev{\rm vev}.\label{three}
\end{eqnarray}
This general picture looks convincingly simple, but its implementation in
realistic models is not. At present, there are several reasons that prevent us
from a complete and satisfactory solution of the mass problem. The quark and
lepton mass spectrum (neglecting neutrinos) spans a range of at least five
orders of magnitude, \ie, from $m_e=0.5\MeV$ to $m_t\gsim 130\GeV$.
If we take as ``normal" the electroweak gauge boson masses, ${\cal O}(80-90)
\GeV$,  then a seemingly ``heavy" top quark ${\cal O}(150\GeV)$ looks perfectly
reasonable, while all other quark and lepton masses look peculiarly small.
Clearly, a natural theory cannot support fundamental Yukawa couplings extending
over five orders of magnitude. The hope has always been \cite{Dimitri} that
several of these Yukawa couplings are {\em naturally zero} at the classical
level, and that quantum corrections generate Yukawa couplings that reproduce
reality. A modern version of this program has arisen in string theory, as we
discuss shortly. We should point out that in a softly broken supersymmetric
theory, several mass parameters arise beyond those in
Eqs.~(\ref{one})--(\ref{three}). However, these lead to ``normal" sparticle
masses, and thus do not relate to the light fermion mass puzzle.

Despite the pessimism expressed above, certain features of the fermion mass
spectrum have been already explored, most
notably in unified theories, where the difference between quark and lepton
masses is attributed to the strong interactions that make the quarks much
heavier than the leptons (of the same generation). In this context, the
successful prediction for the $m_b/m_\tau$ ratio \cite{BEGN} led to the highly
correlated prediction of $N_f = 3$, which was spectacularly confirmed at LEP:
$N_f = 2.980 \pm 0.027$ \cite{LEPC}. An important feature of supergravity
unified models is their ability to trigger radiative spontaneous breaking of
the electroweak symmetry \cite{EWx,LN}, thus explaining naturally why
$m_W/m_{Pl}\approx 10^{-16}$. However, this mechanism only works when the
theory contains a Yukawa coupling of the order of ``$g$", \ie, $y={\cal O}(g)$,
which is naturally identified with the top-quark Yukawa coupling.  In other
words, in supergravity models, a ``heavy" top quark is not only natural, but
it is also needed if we want to have a dynamical understanding of electroweak
symmetry breaking. Finally, string theory -- more precisely its infrared limit,
which  naturally encompasses supergravity -- is characterized by two features
of relevance to us here (see \eg, Refs.~\cite{revamp,fyuks,LNY,SMyuks}):
\begin{enumerate}
\item Most of the Yukawa couplings are naturally zero at the lowest order, and
acquire non-vanishing values progressively at higher orders (through effective
``non-renormalizable" terms), consistent with the spectrum of fermion masses
observed in Nature.
\item Non-zero Yukawa couplings, at lowest order, are automatically of ${\cal
O}(g)$.
\end{enumerate}
Once more, in string theory a ``heavy" top quark is a natural possibility and,
for the first time, we may even have a dynamical explanation for the origin of
its large Yukawa coupling, \ie, ${\cal O}(g)$. We should remark that large
values of the top-quark Yukawa coupling at very high energies have long been
advocated as the explanation for a ``heavy" top quark in connection with the
infrared quasi fixed point of the corresponding renormalization group equation
\cite{nonpert,others,DL,BBO}. However, the origin of such large values has been
usually regarded as a  remnant of new non-perturbative physics at very high
energies \cite{nonpert}, or simply left unspecified. In this note we emphasize
that string theory provides a natural underlying structure where the
experimentally favored values of the top-quark mass can be understood.

\section{SU(5)xU(1) Supergravity: bottom-up view}
Here we briefly describe the most salient features of string-inspired
$SU(5)\times U(1)$ supergravity \cite{EriceDec92}, which constitutes our
bottom-up approach to the prediction for $m_t$. The $SU(5)\times U(1)$ gauge
group (also known as ``flipped $SU(5)$") can be argued to be the simplest
unified gauge extension of the Standard Model. It is unified because the two
non-abelian gauge couplings of the Standard Model ($\alpha_2$ and $\alpha_3$)
are unified into the $SU(5)$ gauge coupling. It is the simplest extension
because this is the smallest unified group which provides neutrino masses. In
this interpretation, minimal $SU(5)$ would appear as a subgroup of $SO(10)$, if
it is to allow for neutrino masses. Moreover, the $SU(5)\times U(1)$ matter
representations entail several simplifications, such as the breaking of the
gauge group via vacuum expectation values of \r{10},\rb{10} Higgs fields, the
natural splitting of the doublet and triplet components of the Higgs pentaplets
and therefore the natural avoidance of dangerous dimension-five proton decay
operators, and the natural appearance of a see-saw mechanism for neutrino
masses.

We supplement the $SU(5)\times U(1)$ gauge group choice with the minimal
matter content which allows it to unify at the string scale
$M_U\sim10^{18}\GeV$, as expected to occur in the string-derived versions
of the model \cite{Lacaze}. This entails a set of intermediate-scale mass
particles: a vector-like quark doublet with mass $m_Q\sim10^{12}\GeV$ and
a vector-like charge $-1/3$ quark singlet with mass $m_D\sim10^6\GeV$
\cite{LNZI}. The model is also implicitly constrained by the requirement
of suitable supersymmetry breaking. We choose two string-inspired scenarios
which have the virtue of yielding universal soft-supersymmetry-breaking
parameters, in contrast with non-universal soft-supersymmetry-breaking
scenarios which occur quite commonly in string constructions
\cite{IL,KL+Ibanez} and may be phenomenologically troublesome \cite{EN}.
These scenarios are examples of the no-scale supergravity framework
\cite{Lahanas+EKNI+II,LN} in which the dimensional parameters of the theory are
undetermined at the classical level, but are fixed by radiative corrections,
thus including the whole theory in the determination of the low-energy
parameters. In the {\em moduli} scenario, supersymmetry breaking is driven
by the vev of the moduli fields ($T$), and gives $m_0=A=0$, whereas in the
{\em dilaton} scenario \cite{KL+Ibanez} supersymmetry breaking is driven
by the vev of the dilaton field ($S$) and entails $m_0={1\over\sqrt{3}}m_{1/2},
A=-m_{1/2}$. Thus, the supersymmetry breaking sector depends on only one
parameter (\ie, $m_{1/2}$).

The procedure to extract the low-energy predictions of the models outlined
above is rather standard (see \eg, Ref. \cite{aspects}): (a) the bottom-quark
and tau-lepton masses, together with the input values of $m_t$ and $\tan\beta$
are used to determine the respective Yukawa couplings at the electroweak scale;
(b) the gauge and Yukawa couplings are then run up to the unification scale
$M_U=10^{18}\GeV$ taking into account the extra vector-like quark doublet
($\sim10^{12}\GeV$) and singlet ($\sim10^6\GeV$) introduced above
\cite{sism,LNZI}; (c) at the unification scale the soft-supersymmetry breaking
parameters are introduced (\ie, moduli and dilaton scenarios) and the scalar
masses are then run down to the electroweak scale; (d) radiative electroweak
symmetry breaking is enforced by minimizing the one-loop effective potential
which depends on the whole mass spectrum, and the values of the Higgs mixing
term $|\mu|$ and the bilinear soft-supersymmetry breaking parameter $B$ are
determined from the minimization conditions; (e) all known phenomenological
constraints on the sparticle and Higgs masses are applied (most importantly the
LEP lower bounds on the chargino and Higgs-boson masses), including the
cosmological requirement of a not-too-young Universe.

The three-dimensional parameter space of this model (\ie, $m_{1/2},\tan\beta$
and the top-quark mass) has been explored in detail in Refs.~\cite{LNZI} and
\cite{LNZII} for the moduli and dilaton scenarios respectively.  More recently,
we have investigated further constraints on the parameter space, including: (i)
the CLEO limits on the $b\to s\gamma$ rate \cite{bsgamma,bsg-eps}, (ii) the
long-standing limit on the anomalous magnetic moment of the muon \cite{g-2},
(iii) the electroweak LEP high-precision measurements in the form of the
$\epsilon_{1},\epsilon_b$ parameters \cite{ewcorr+eps1-epsb,bsg-eps}, (iv) the
non-observation of anomalous muon fluxes in underground detectors (``neutrino
telescopes") \cite{NT}, and (v) the
possible constraints from trilepton searches at the Tevatron \cite{LNWZ}.

\section{SU(5)xU(1) Supergravity: top-down view}
In the context of string model-building, the $SU(5)\times U(1)$ structure
becomes even more important, since the traditional grand unified gauge groups
($SU(5),SO(10),E_6$) cannot be broken down to the Standard Model gauge group in
the simplest (and to date almost unique) string  constructions, because of the
absence of adjoint Higgs representations \cite{ELN}. This reasoning is not
applicable to the $SU(5)\times U(1)$ gauge group, since the required
\r{10},\rb{10} representations are very common in string model building
\cite{revamp,LNY}. As a ``descendant" of string theory, $SU(5)\times U(1)$
supergravity is characterized by two basic features: (a) a large top-quark
Yukawa coupling: ${\cal O}(g)$, and (b) the no-scale structure. Notice that (b)
in conjunction with (a), not only triggers radiative electroweak breaking but,
in principle, may also determine {\em dynamically} the magnitude of the
supersymmetry breaking scale \cite{Lahanas+EKNI+II,LN}. As mentioned above,
string unification occurs at the scale $M_U\sim10^{18}\GeV$ \cite{Lacaze}, and
this has been seen to occur in explicit $SU(5)\times U(1)$ string models
\cite{LNY}.

Of more relevance to the present discussion is the composition of the Yukawa
sector in $SU(5)\times U(1)$ string models. The usual situation
\cite{revamp,fyuks,SMyuks} is that at the cubic level of superpotential
interactions, string symmetries allow only few couplings among the matter
fields containing the quarks, leptons, and Higgs bosons of the low-energy
theory. A particularly simple solution to the question of how to assign
low-energy fields to the string representations consists of having only the
top-quark, bottom-quark, and tau-lepton Yukawa couplings be non-vanishing.
Further assumptions lead to a scenario with
$\lambda_t=\lambda_b=\lambda_\tau=\sqrt{2}g$ at the string scale, where $g$ is
the unified gauge coupling determined by the
vacuum expectation value of the dilaton field in the top-down approach, or
by the unification condition in the bottom-up approach. This is however not a
robust prediction since various unknown mixing angles could possibly destroy
this relation. Moreover, it is possible that the bottom-quark and tau-lepton
Yukawa couplings could be suppressed relative to the top-quark Yukawa coupling
\cite{SMyuks}. What is a robust prediction is the magnitude of the top-quark
Yukawa coupling
\beq
\lambda_t(M_U)=\sqrt{2}g\cos\theta_t\,, \label{topyuk}
\eeq
where $\cos\theta_t$ is a possible mixing angle factor. The bottom-quark and
tau-lepton Yukawa couplings are not necessarily equal at the string scale,
since no obvious symmetry principle is at play in $SU(5)\times U(1)$ (as
opposed to the case of $SU(5)$). Nonetheless, equality of these Yukawa
couplings does occur in many explicit $SU(5)\times U(1)$ string models
\cite{revamp,fyuks,LNY}. The Yukawa couplings for the first- and
second-generation quarks and leptons appear at the quartic or higher
non-renormalizable order \cite{KLN,fyuks} and are naturally suppressed
relative to the cubic level Yukawa couplings, in agreement with the observed
hierarchical mass spectrum.

\section{The Yukawa sector and the value of $m_t$}
{}From the bottom-up approach we are able to compute the value of the
third-generation Yukawa couplings at the string scale in terms of $m_t$ and
$\tan\beta$. (These string-scale Yukawa couplings also depend on $m_b$ and
$\alpha_3(M_Z)$.) In Fig.~\ref{yukt} we show the top-quark Yukawa coupling at
the string scale versus the top-quark mass for various values of $\tan\beta$.
As expected, a Landau pole is encountered in the running of the Yukawa coupling
if the top-quark mass exceeds a maximum value at low energies. For example,
$m_t\lsim170\GeV$ is required for $\tan\beta=2$. Values of $\tan\beta$ larger
than those shown are indistinguishable from the $\tan\beta=10$ curve. The
dependence on $\alpha_3(M_Z)$ and $m_b$ is rather small in this case, \ie,
comparable to the thickness of the lines for $\alpha_3(M_Z)=0.118\pm0.007$ and
$m_b=4.25-4.9\GeV$.

The above are the results of the bottom-up approach.
On the other hand, from the top-down approach we expect values of $\lambda_t$
as given in Eq.~(\ref{topyuk}), which are shown as dashed lines on
Fig.~\ref{yukt} for two typical cases. Here $g\approx0.84$ is obtained from
the running of the gauge couplings up to the string scale. These values of
$\lambda_t$ do not exceed the unitarity requirement of Ref.~\cite{DL}
($\lambda_t<4.8$) or the perturbative criterion of Ref.~\cite{BBO}
($\lambda_t<3.3$). Thus, the experimentally preferred top-quark masses
(direct Tevatron limits $m_t>131\GeV$ \cite{D0top} and indirect fits to
the electroweak data $m_t\approx140\pm20\GeV$ \cite{Altarelli}) can be
naturally understood in string models, and do not require the existence of new
non-perturbative interactions at the unification scale.

{}From the bottom-up approach we also obtain the values for the bottom-quark
and
tau-lepton Yukawa couplings at the string scale, as shown in Figs.~\ref{yukb}
and \ref{yukbtau} (for $\alpha_3(M_Z)=0.118$). In Fig.~\ref{yukb}, the
bottom-quark Yukawa coupling is plotted against the top-quark Yukawa coupling
for various values of $\tan\beta$ (2,6,10,20). Along the (solid) lines the
top-quark mass varies as shown. The two sets of curves for each value of
$\tan\beta$ correspond to the representative choices of $m_b=4.25$ and
$4.9\GeV$. The dashed lines for $\tan\beta=10$ show the decrease in
$\lambda_b(M_U)$ due to a shift in $\alpha_3(M_Z)$ from 0.118 to 0.125. The
corresponding shifts for larger (smaller) values of $\tan\beta$ are
proportionally larger (smaller). Values of $\tan\beta$ larger than the ones
shown, when allowed by the theoretical constraints on the model, simply yield
proportionally larger values of $\lambda_b(M_U)$.

In Fig.~\ref{yukbtau} the bottom-quark Yukawa coupling is plotted against
the tau-lepton Yukawa coupling for various values of $\tan\beta$ (2,6,10,20).
Two representative values of $m_b$ have been chosen (4.25 and 4.9 GeV) which
are only visibly distinguished for $\tan\beta=20$, as indicated. Also,
$\alpha_3(M_Z)=0.118$ has been chosen. Along the vertical lines the top-quark
mass increases from bottom to top. The effect of shifts in $\alpha_3(M_Z)$ is
to extend the vertical lines slightly. It is interesting to note that the
traditional $\lambda_b=\lambda_\tau$ relation (as would be required in an
$SU(5)$ model) can be obtained for the largest values of $m_t$ and for the
larger values of $m_b$. However, the range
\beq
0.7\lsim\lambda_b/\lambda_\tau\lsim1
\eeq
is a more realistic estimate of what would be required from a string model
in the top-down approach. Such deviations from the $\lambda_b=\lambda_\tau$
relation have been explored in the literature \cite{BBO} and have been shown to
weaken significantly the tight constraint on the $(m_t,\tan\beta)$ plane which
otherwise results from imposing the $\lambda_b=\lambda_\tau$ relation.

As discussed above, the allowable free parameters are reduced to a minimal
number in $SU(5)\times U(1)$ supergravity, allowing severe experimental
scrutiny. An interesting exercise along these lines consists of determining
the size of the allowed parameter space (in the $(m_{1/2},\tan\beta)$ plane)
as a function of $m_t$, hoping that the correlations among the model variables
and their intricate dependence on $m_t$ may show a preference for particular
values of the top-quark mass. The results of this exercise, when only the
basic theoretical and experimental LEP constraints are imposed, are shown in
Fig.~\ref{counts} (``theory+LEP" curves). The drop in the curves near
$m_t=190\GeV$ has been studied in detail (for $m_t=180,185,187,188,189\GeV$)
and corresponds to encountering a Landau pole in the top-quark Yukawa coupling
below the string scale \cite{DL,aspects}. Imposing in addition all of the
direct and indirect experimental constraints mentioned above (\ie, $b\to
s\gamma$, $(g-2)_\mu$, neutrino telescopes, and $\epsilon_{1,b}$)
we obtain the curves labelled ``ALL" in Fig.~\ref{counts} \cite{Easpects}.
These curves still do not show any obvious preference for particular values
of $m_t$. However, $m_t\lsim180\GeV$ is now required, basically to fit the
precise LEP electroweak data \cite{Easpects}. This exercise is rather
interesting and should be repeated as present experimental constraints are
tightened or new constraints arise.

\section{Conclusions}
We have shown that the currently experimentally preferred values of the
top-quark mass are naturally understood in a top-down approach in the context
of string models. We have studied this point explicitly in the context of
$SU(5)\times U(1)$ supergravity, which is to be viewed as the bottom-up
approach to physics at the string scale. Using the bottom-up approach we have
also found that the ratio of the bottom-quark to tau-lepton Yukawa couplings at
the string scale is required to be in the range
$0.7\lsim\lambda_b/\lambda_\tau\lsim1$, depending on the values of $m_t$ and
$m_b$. This result is consistent with $SU(5)\times U(1)$ symmetry, which
does {\em not} require the equality of these Yukawa couplings in the unbroken
symmetry phase of the theory. Finally, as a means of possibly predicting the
value of $m_t$, we have proposed a procedure whereby the size of the allowed
parameter space is determined as a function of $m_t$. Since all sparticle and
Higgs-boson masses and couplings, and therefore all observables calculated
from them, depend non-trivially on $m_t$ (mostly through the radiative breaking
mechanism), such procedure could show a preference towards particular values of
$m_t$. At present no such preference is clearly observed, except for the
high-precision LEP data requirement of $m_t\lsim180\GeV$. Nonetheless, future
more sensitive experimental constraints may produce more clear effects.
This present relative insensitivity to the value of $m_t$ should not obscure
the fact that all experimentally preferred values of $m_t$ are allowed in
$SU(5)\times U(1)$ supergravity, even after the many theoretical and
experimental constraints have been applied to the model. We should remark that
this procedure could also be applied to more general classes of
supergravity models, which have been recently studied in the literature
\cite{recent}, as a means of gauging the experimental viability of these
models.

\section*{Acknowledgements}
This work has been supported in part by DOE grant DE-FG05-91-ER-40633.

\newpage

\section*{Figure Captions}
\begin{enumerate}
\item The top-quark Yukawa coupling at the string scale in $SU(5)\times U(1)$
supergravity versus the top-quark mass for fixed values of $\tan\beta$ (larger
values of $\tan\beta$ overlap with the $\tan\beta=10$ curve). The dashed lines
indicate typical string-like predictions for the Yukawa coupling.
\label{yukt}
\item The bottom-quark Yukawa coupling versus the top-quark Yukawa coupling at
the string scale in $SU(5)\times U(1)$ supergravity for various values of
$\tan\beta$ (2,6,10,20), two values of $m_b$ (4.25 and 4.9 GeV), and
$\alpha_3(M_Z)=0.118$. The top-quark mass varies along the curves as indicated.
The dashed lines for $\tan\beta=10$ show the effect of varying $\alpha_3(M_Z)$
from 0.118 to 0.125. The magnitude of this effect scales with $\tan\beta$.
\label{yukb}
\item The bottom-quark Yukawa coupling versus the tau-lepton Yukawa coupling at
the string scale in $SU(5)\times U(1)$ supergravity for various values of
$\tan\beta$ (2,6,10,20), two values of $m_b$ (4.25 and 4.9 GeV), and
$\alpha_3(M_Z)=0.118$. The value of $m_t$ increases from bottom to top along
the vertical lines. Note that $0.7\lsim\lambda_b/\lambda_\tau\lsim1$ is
obtained.
\label{yukbtau}
\item The number of allowed points in parameter space of no-scale $SU(5)\times
U(1)$ supergravity in the moduli and dilaton scenarios, as a function of $m_t$
when the basic theoretical and experimental LEP constraints have been imposed
(``theory+LEP"), and when all known direct and indirect experimental
constraints have been additionally imposed (``ALL"). Note that
$m_t\lsim180\GeV$ is required.
\label{counts}
\end{enumerate}

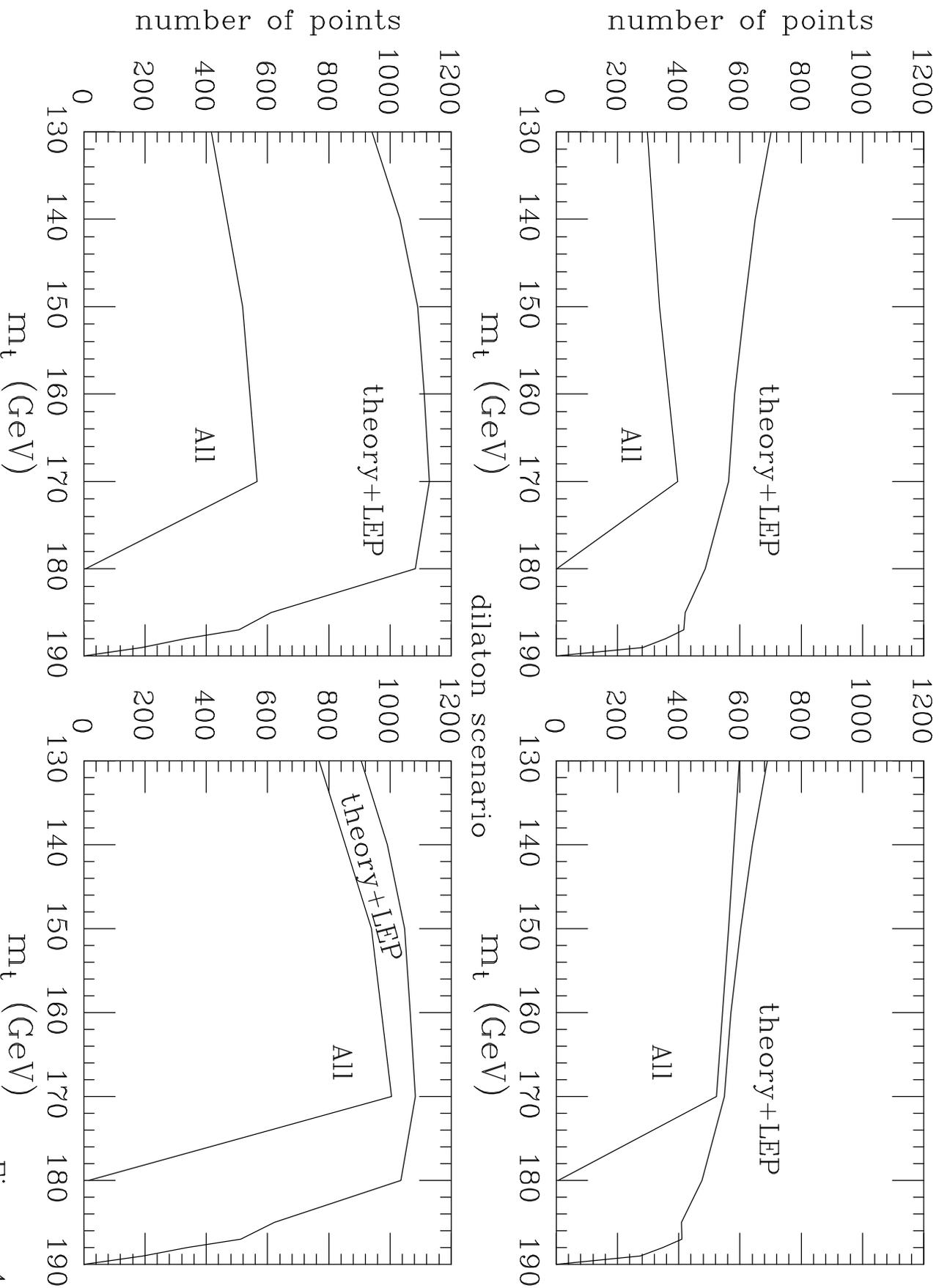
\end{document}